\documentclass{article}

\usepackage{arxiv}

\usepackage[utf8]{inputenc} % allow utf-8 input
\usepackage[T1]{fontenc}    % use 8-bit T1 fonts
\usepackage{hyperref}       % hyperlinks
\usepackage{url}            % simple URL typesetting
\usepackage{booktabs}       % professional-quality tables
\usepackage{amsfonts}       % blackboard math symbols
\usepackage{nicefrac}       % compact symbols for 1/2, etc.
\usepackage{microtype}      % microtypography
\usepackage{lipsum}		% Can be removed after putting your text content
\usepackage{graphicx}
\usepackage{natbib}
\usepackage{doi}
\usepackage{latexsym}
\usepackage{graphicx}
\usepackage{mathptmx}
\usepackage[T1]{fontenc}
\usepackage{subcaption}
\usepackage{listings}
\usepackage{xcolor}
\usepackage{makecell}
\usepackage{pifont}
\usepackage{amsmath}
\usepackage{amsfonts}
\usepackage{amssymb}
\usepackage{amsbsy}
\usepackage{amsthm}
\usepackage{appendix}
\lstdefinestyle{json}{
  basicstyle=\ttfamily\footnotesize,
  columns=fullflexible,
  breaklines=true,
  breakatwhitespace=false,
  showstringspaces=false,
  frame=single,
  numbers=left,
  numberstyle=\tiny\color{black!50},
  numbersep=8pt,
  tabsize=1,
    literate=
    {①}{{(i)}}1
    {②}{{(ii)}}1
    {③}{{(iii)}}1
    {±}{{$\pm$}}1
    {—}{{-}}1
    {≥}{{$\geq$}}1
}

\title{Ambiguity Detection and Elimination in Automated Executable Process Modeling}

\author{
    {\hspace{1mm}Ion Matei} \\
	Fujitsu Research of America\\
	\texttt{imatei@fujitsu.com} \\
	\And
	{\hspace{1mm}Praveen Kumar Menaka Sekar} \\
	University of Maryland, College Park\\
	\texttt{praveenm@umd.edu} \\
    \And    
    {\hspace{1mm}Maksym Zhenirovskyy} \\
	Fujitsu Research of America\\
	\texttt{mzhenirovskyy@fujitsu.com} \\
    \And
    {\hspace{1mm}Hon Yung Wong} \\
	Fujitsu Research of America\\
	\texttt{awong@fujitsu.com} \\
    \And
    {\hspace{1mm}Sayuri Kohmura} \\
	Fujitsu Limited\\
	\texttt{kohmura.sayuri@fujitsu.com} \\   
    \And
    {\hspace{1mm}Shinji Hotta} \\
	Fujitsu Limited\\
	\texttt{hotta\_s@fujitsu.com} \\       
    \And
    {\hspace{1mm}Akihiro Inomata} \\
	Fujitsu Limited\\
	\texttt{akiino@fujitsu.com} \\        
}

\hypersetup{
pdftitle={A template for the arxiv style},
pdfsubject={q-bio.NC, q-bio.QM},
pdfauthor={David S.~Hippocampus, Elias D.~Striatum},
pdfkeywords={First keyword, Second keyword, More},
}

\begin{document}
\maketitle

\begin{abstract}
Automated generation of executable Business Process Model and Notation (BPMN) models from natural-language specifications is increasingly enabled by large language models. However, ambiguous or underspecified text can yield structurally valid models with different simulated behavior. Our goal is not to prove that one generated BPMN model is semantically correct, but to detect when a natural-language specification fails to support a stable executable interpretation under repeated generation and simulation. We present a diagnosis-driven framework that detects behavioral inconsistency from the empirical distribution of key performance indicators (KPIs), localizes divergence to gateway logic using model-based diagnosis, maps that logic back to verbatim narrative segments, and repairs the source text through evidence-based refinement. Experiments on diabetic nephropathy health-guidance policies show that the method reduces variability in regenerated model behavior. The result is a closed-loop approach for validating and repairing executable process specifications in the absence of ground-truth BPMN models.
\end{abstract}

% keywords can be removed

\section{Introduction}

Natural-language policies and clinical guidelines often need to be converted into executable models to support simulation, quantitative evaluation, and data-driven decision making. In healthcare and public policy, this matters because stakeholders increasingly want to assess intervention logic, estimate key performance indicators (KPIs), and compare alternative strategies before deployment. BPMN provides a useful representation because it captures workflow logic in a formal and executable form while remaining accessible to domain experts~\cite{omg_bpmn}. Recent advances in large language models (LLMs) have made it possible to automatically generate BPMN models from policy documents and clinical narratives~\cite{kourani2024processmodelingllm,matei2026syscon,vanDerAa2019text2process}. However, automatic generation alone does not solve the validation problem. Natural-language specifications are often ambiguous, incomplete, or context-dependent~\cite{iso29148_2018,berry2003ambiguity}. As a result, repeated BPMN generation from the same source text can produce multiple syntactically valid and executable models with different simulated behavior. Structural checks such as soundness, gateway matching, and deadlock freedom remain necessary~\cite{corradini2020correctness}, but they are not sufficient here. A model may be structurally correct and still encode a different interpretation of the source narrative than intended. For automatically generated executable models, the central question is whether the source text supports a \emph{stable executable interpretation}.

This paper studies that question in a setting where no ground-truth BPMN model is available. Our goal is not to prove that one generated BPMN model is semantically correct, but to detect when a natural-language specification fails to support a stable executable interpretation under repeated generation and simulation. We do this through behavioral evidence: we generate multiple BPMN models from the same narrative, simulate them on the same input data, and analyze the empirical distribution of KPI outcomes. A highly dispersed distribution indicates that the source specification supports competing executable interpretations under the generation pipeline. In this setting, generation consistency is used as a proxy for reliability, not as a proof of semantic correctness. We recently developed an end-to-end LLM-driven pipeline for generating executable BPMN models from healthcare policy documents and evaluating them through simulation~\cite{praveen_git,sekar2026automaticgenerationexecutablebpmn}. The resulting models are further optimized in downstream applications~\cite{shimaoka2026structureawareoptimizationdecisiondiagrams}. In that system, we observed a recurrent failure mode: some policy descriptions produced multiple coherent but behaviorally different models, even though the models were structurally valid. This divergence appeared as elevated entropy and multimodal KPI distributions. Because the effect did not appear uniformly across all policies, it suggested that the source text was an important contributor to the observed instability. In this paper, we define \emph{ambiguity} as a property of a natural-language specification that admits multiple logically coherent executable interpretations with different simulated behavior. These observations motivate the following research question: \textit{Given behaviorally inconsistent executable models generated from the same narrative, how can we localize the source of divergence in the underlying text and revise the specification so that it yields more consistent executable behavior?}

To answer this question, we make three main contributions. First, we introduce a simulation-based ambiguity detection method that uses repeated BPMN generation, KPI-output distributions, and normalized entropy to identify when a specification fails to support a stable executable interpretation. Second, we propose a diagnosis-driven localization procedure that selects representative models from dominant KPI classes and applies model-based diagnosis (MBD)~\cite{deKleer1987diagnosis,reiter1987theory} to link divergent simulated behavior to gateway-level process logic and then to verbatim narrative segments in the source text. Third, we develop an evidence-based repair loop that rewrites only the affected narrative segments using authoritative supporting material and validates the repair through regeneration and re-simulation, yielding improved behavioral consistency. We demonstrate the method on diabetic nephropathy~\cite{10.3389/fendo.2023.1195167} health-guidance policies from two Japanese municipalities. These case studies contain nested eligibility rules, operational branching, and implicit logical dependencies that are difficult to translate consistently into executable models. The results show that repairing the identified ambiguities leads to more concentrated KPI distributions and substantially more stable regenerated BPMN behavior.
% The rest of the paper is organized as follows. Section~\ref{sec:Automatic Detection} presents the detection, diagnosis, localization, and repair framework. Section~\ref{sec:Experimental Results} reports the case-study results. Section~\ref{sec:Comparison} compares the proposed approach with prior work on ambiguity detection, repair, and automated process modeling.

\section{Ambiguity Detection and Repairs}
\label{sec:Automatic Detection}

This section presents our framework for detecting and repairing ambiguity in natural-language process descriptions using executable BPMN evidence. %Behavioral divergence is first identified through distribution-level simulation analysis and then localized through model-based diagnosis. The diagnosed divergence is mapped back to the source narrative through LLM-based analysis and resolved through minimal, traceable refinements validated by regeneration and re-simulation.

\subsection{Automatic Text-to-Executable Model Generation}
\label{sec:Automatic Generation}
We build on our prior pipeline for generating executable BPMN models from natural-language clinical specifications~\cite{sekar2026automaticgenerationexecutablebpmn}. The pipeline preprocesses the source document, uses an LLM to extract tasks, events, gateways, and data dependencies, generates BPMN XML, executes the model with a workflow engine~\cite{spiffworkflow}, and aggregates execution traces into policy KPIs. For the present paper, the critical property is that gateway conditions are grounded in available input data variables and KPI-producing activities are explicitly and automatically mapped, so repeated generations can be compared behaviorally under the same simulated population.

\subsection{BPMN Simulation Output Distribution Analysis}
To assess macro-scale behavioral uncertainty in BPMN generation, we analyze the distribution of simulated KPI outputs using normalized entropy. Each generated BPMN model is simulated over the same synthetic population, and the resulting aggregate KPI vector is treated as one sample,
\(
y = (y_1, y_2, \dots, y_d),
\)
with $d$ denoting the number of KPIs. From these samples, we construct an empirical distribution over the distinct KPI output vectors observed across generated models. Because the KPI outputs are sparse, we represent this distribution as a discrete probability mass function over the set of unique output combinations. Let $\mathcal{Y} = \{y^{(1)}, \dots, y^{(K)}\}$ denote the set of unique KPI combinations, and let $p_i$ be the empirical probability of observing $y^{(i)}$. We then compute the normalized Shannon entropy of this distribution as
\(
H_{\text{norm}} =
-\left(\sum_{i=1}^{K} p_i \log_2 p_i\right)/\log_2 |\mathcal{Y}|,
\)
which takes values in $[0,1]$ and provides a scale-independent measure of output dispersion. 
We use $H_{\text{norm}}$ as a heuristic measure of generation consistency: lower values indicate concentration around one or a few KPI combinations, whereas higher values indicate multiple competing executable interpretations. In the experiments, we refer to four qualitative ranges certainty: \textit{very high} ($\leq 0.30$), \textit{high} ($(0.30,0.50]$), \textit{moderate} ($(0.50,0.70]$), and \textit{low} ($>0.70$) consistency.

\subsection{Model-Based Diagnosis for Process Model Validation}

When two BPMN models represent different interpretations of the same text, their equivalence must be assessed behaviorally rather than only structurally. In our setting, this is done through simulation. If the models produce different KPI outputs for the same inputs, we must identify which model elements explain that difference. We formulate this as a diagnosis problem and use model-based diagnosis (MBD)~\cite{deKleer1987diagnosis,reiter1987theory}. We treat gateways as potentially faulty components because they encode the main decision logic and are the most likely source of divergent behavior. Prior work compares BPMN models through event-structure relations such as causality, conflict, and repetition~\cite{10.1016/j.is.2015.09.009}. Our goal is different: we use diagnosis inside an ambiguity detection and repair loop, so direct diagnosis of gateway decisions from execution outcomes is more appropriate.

From the empirical distribution of simulation results, we select two representative models from two dominant KPI classes. One is designated as the reference and the other as the target for diagnosis, with no assumption on correctness. Instead, we choose the diagnosis direction that produces the smaller minimum-diagnosis set, since this gives a more localized explanation. In our formulation, the system description ($SD$) consists of the target-model process structure together with its gateway logic. The component set ($COMPS$) is the set of gateways in the target model. The observations ($OBS$) are activity-level KPI outputs obtained by comparing the simulations of the reference and target models. An observation is marked as discrepant when the reference and target outputs differ. Formally,
\(
OBS_{\text{disc}} = \{o \in OBS \mid o_{\text{ref}} \neq o_{\text{tgt}} \}.
\)
These discrepant observations serve as symptoms in the diagnostic process.

To localize the source of error, we compare the reference and target models only on input subsets for which their outputs differ. Agreement on other inputs indicates conditional inconsistency, so each divergent subset defines a separate diagnostic context. For each execution trace $\tau$ in such a context, we construct an ordered sequence
\(
S(\tau) = \langle (t_1, K_1), \ldots, (t_n, K_n) \rangle,
\)
where each pair contains a task and the KPI it produces. We include only KPI-producing tasks, and when multiple KPIs are associated with the same activity, their names are sorted deterministically. We then compare the reference and target sequences and identify the first point at which they diverge. The divergence may appear as a missing output, an extra output, or an incorrect output. Outputs after the first divergence are treated as downstream consequences and are not used to construct conflicts. Let $t_{\text{last}}$ be the last correct task and $t_{\text{first}}$ the first erroneous task for a given divergent trace pair. Here, the ordering relation is the execution order along the target trace under the corresponding input case. We define the corresponding conflict set as
\(
CONF_i = \{g \in COMPS \mid t_{\text{last}} < g < t_{\text{first}}\}.
\)
Thus, each divergent input-output case yields one or more conflict sets containing only the target-model gateways between the last correct and first incorrect outputs. This positional restriction reduces conflict size. The full diagnosis problem is built from the collection of conflict sets $\mathcal{C}$ obtained across all divergent subsets. We then compute minimal diagnoses as minimal hitting sets over $\mathcal{C}$, following standard MBD theory. Although hitting-set computation is NP-hard in general, the bounded size of the conflict sets makes enumeration practical in our setting.

\textit{Diagnosis Refinement through Path-Based Analysis:}
We further refine diagnoses using input-conditioned equivalence of exercised gateway decision functions rather than structural similarity between BPMN models. A target gateway is removed from a diagnosis if its normalized logical condition is identical to a condition exercised in the reference trace for the same input cases. In that case, the gateway does not explain the observed output difference. This equivalence check is implemented through comparison of normalized abstract syntax trees (ASTs)~\cite{aho2006compilers}.

\subsection{Ambiguity Elimination and Narrative Refinement}

The repair stage revises only the text segments identified in the ambiguity report. It uses the original narrative, the structured ambiguity report, and authoritative supplemental material to select one supported interpretation and rewrite the affected passage with minimal change. The result is a repaired specification plus traceability metadata linking each revision to the original ambiguity and its supporting evidence. The ambiguity elimination prompt implements a four-step procedure.

\textit{Ambiguity localization and mapping:} Each ambiguity instance is mapped to its exact location in the source narrative. This constrains the revision process to the affected text and avoids unnecessary edits elsewhere in the document.

\textit{Evidence-based interpretation selection:} The competing interpretations identified in the ambiguity report are evaluated against the supplemental material. The selected interpretation must be justified by explicit supporting excerpts. Unsupported assumptions are not allowed.

\textit{Minimal disambiguation synthesis:} The ambiguous text is then rewritten to make the intended logic explicit. This includes clarifying logical operators such as \texttt{AND} and \texttt{OR}, making temporal dependencies explicit, and resolving underspecified conditions. The revision is kept as small as possible so that the original terminology and phrasing are preserved while the identified competing interpretations are reduced to a single intended executable reading.

\textit{Narrative reconstruction:} The revised passages are reinserted into the full narrative to produce the repaired specification.

The output of this stage is a structured ambiguity elimination report that contains the revised narrative together with traceability metadata, including ambiguity identifiers, rewritten excerpts, justifications, and the supporting evidence used for each revision.

\section{Experimental Results}
\label{sec:Experimental Results}

As part of an ongoing policy digitalization project, we automated the generation of simulation-ready BPMN models for diabetic nephropathy policies. During experimentation, we observed that some policy descriptions produced substantial variability in model outputs. Because no ground-truth BPMN models were available, we used simulation consistency as a proxy for generation reliability. We quantified this consistency using normalized entropy over the KPI distributions. This section reports results for two policies that exhibited high variability and shows how our method detects and repairs the ambiguities that give rise to that variability. Our case study uses diabetic nephropathy~\cite{10.3389/fendo.2023.1195167} health-guidance policies from two Japanese municipalities. These policies contain multi-stage screening rules, nested decisions, and implicit temporal dependencies. The automated pipeline, including PDF extraction and GPT-5.1-based translation, could itself introduce ambiguity. However, the translations were reviewed by native speakers and were therefore considered unlikely to be the main source of the observed model-generation and simulation variability. The prompts for the ambiguity detection and repair, together with the input data, ambiguity detection and repair reports, representative models and a tool demo movie can be found at \url{https://github.com/ionmatei/ambiguity-detection}. All steps that involve an LLM use GPT-5.1.

\subsection{Experimental Setup}

The generation pipeline described in Section~\ref{sec:Automatic Generation} converts policy documents into executable BPMN models and evaluates them using five KPIs: \emph{Notification Count} (NC), \emph{Health Guidance Count} (HC), \emph{Guidance Resource Utilization} (RU), \emph{Health Improvement Rate} (HI), and \emph{Medical Cost Savings} (CS). These KPIs are derived from the Program for Preventing the Progression of Diabetic Nephropathy~\cite{JMA2024Nephropathy}. NC counts executions of notification tasks. HC counts patients who receive guidance. RU measures how heavily guidance activities use available capacity and penalizes overload. HI estimates the fraction of patients who achieve clinically meaningful improvement, such as HbA1c reduction, using evidence-based response rates. CS estimates avoided treatment costs due to reduced disease progression over a fixed time horizon.

For each policy, we independently generate 100 BPMN models, simulate each model on the same synthetic patient population, and aggregate five population-level KPIs. We then estimate the empirical KPI-output distribution and compute its normalized entropy. Policies with high entropy are passed to the ambiguity-repair pipeline: MBD localizes the divergent decision constructs, ambiguity localization links them to source text, and evidence-based rewriting produces a repaired specification for regeneration and re-simulation.

In the diagnosis stage, we select a reference model and a target model from dominant model classes associated with different KPI combinations. The reference model serves as the diagnostic baseline, and the target model is diagnosed relative to it. The prompt explicitly states that ``reference'' and ``target'' do not imply that one model is correct and the other is incorrect. We choose the direction of diagnosis based on which assignment produces the smaller minimum-diagnosis set, as this yields a more localized and interpretable diagnostic explanation. The authoritative supplemental material used in the repair stage is a Tokyo Program for Prevention of Severe Progression of Diabetic Nephropathy~\cite{tokyo_diabetic_nephropathy_program_2022}, with the relevant translated paragraphs used for repairs found at (\url{https://github.com/ionmatei/ambiguity-detection/blob/main/supplemental_material.pdf}).

\subsection{City 1}
\begin{figure}[htp!]
    \centering
    \includegraphics[width=0.7\textwidth]{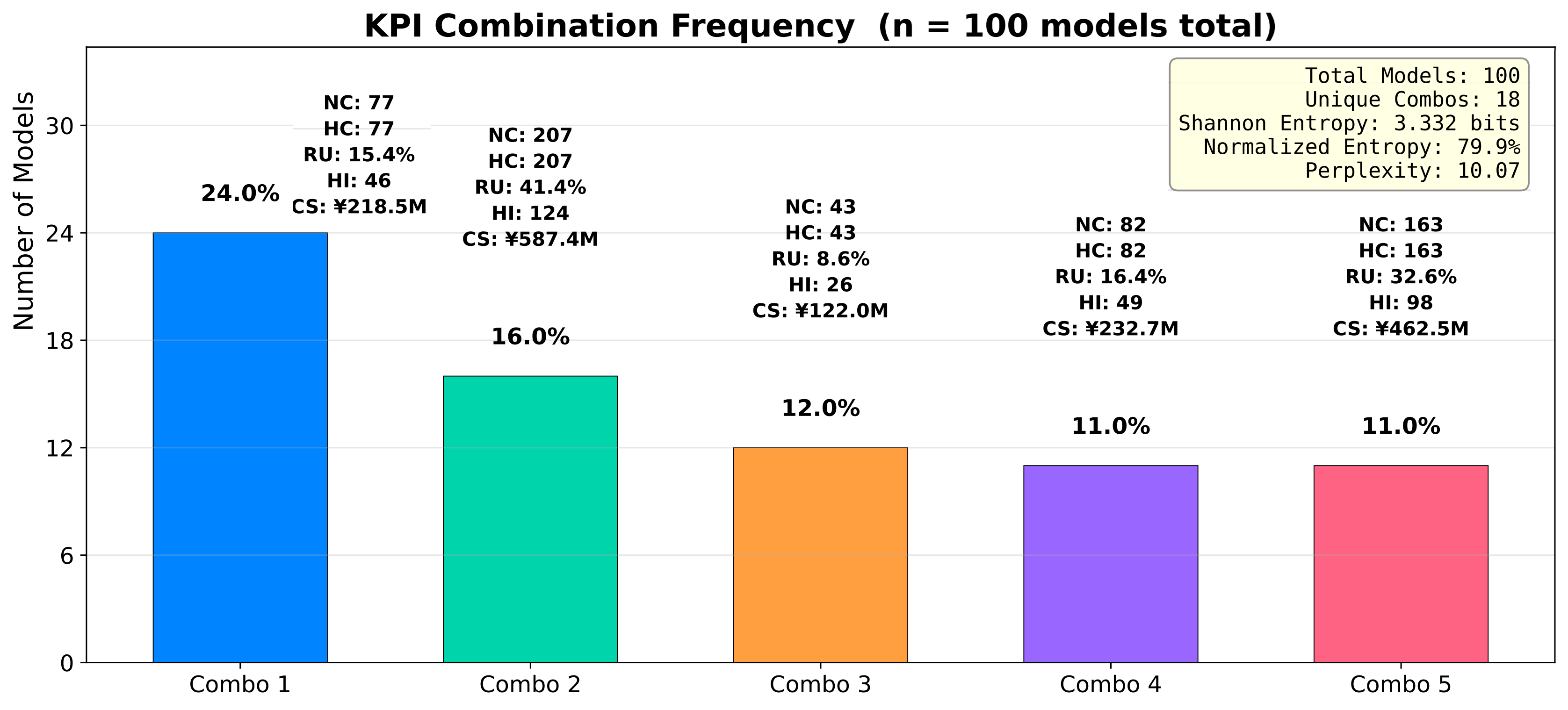}
    \caption{City 1, original process description --- Distribution of all five KPIs across 100 generated models, showing the five most frequent KPI combinations.}
    \label{fig:city_1_KPI_uncertainty_original}
\end{figure}

The extracted City 1 policy document can be found at \url{https://github.com/ionmatei/ambiguity-detection/blob/main/city-1/city_1_raw_policy.pdf}. The policy mainly defines an administrative framework for long-term project management and outcome tracking. It specifies evaluation targets, such as HbA1c and eGFR improvement rates over a six-year horizon, and describes an organizational structure involving medical associations and private operators. Participation is driven through broad publicity and recommendation notices. Using our automated BPMN generation pipeline, we generated 100 models and simulated them on synthetically generated patient data \url{https://github.com/ionmatei/ambiguity-detection/blob/main/input-data/test_data.csv}. Figure~\ref{fig:city_1_KPI_uncertainty_original} shows the resulting KPI histogram. The distribution contains several distinct KPI combinations, indicating substantial variability across generated models. This suggests that the policy description admits multiple executable interpretations under our generation pipeline.

\begin{figure}[htp!]
    \centering
    \includegraphics[width=0.67\textwidth]{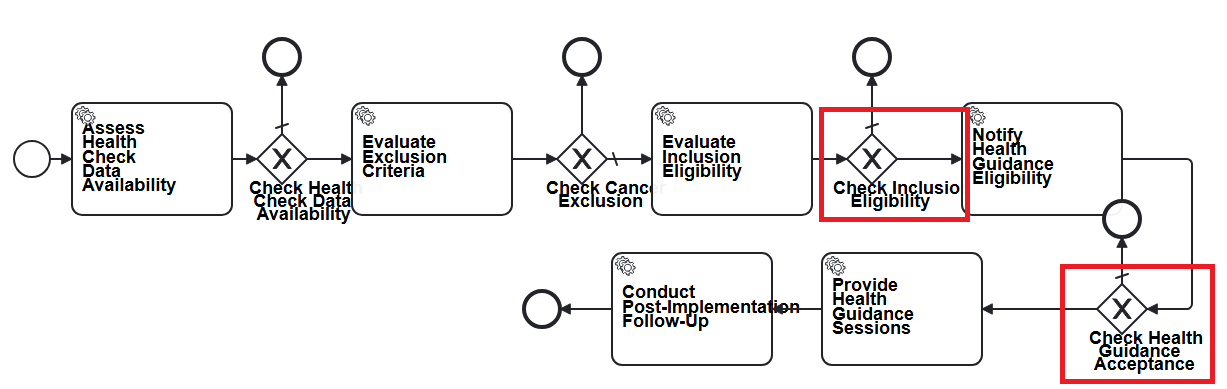}
    \caption{City 1 --- Target models (minimum diagnosis set highlighted in red).}
    \label{fig:city_1_target_model}
\end{figure}

\begin{figure}[htp!]
    \centering
    \includegraphics[width=0.67\textwidth]{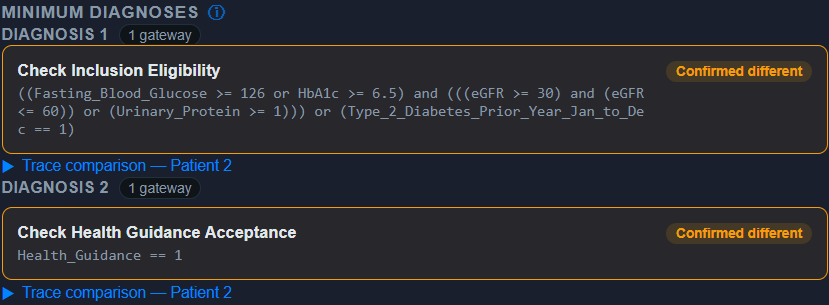}
    \caption{City 1 --- Minimum diagnosis set identifying target-model components that explain the simulation differences.}
    \label{fig:city_1_minimum_diagnosis}
\end{figure}

Next, we select two models for diagnosis: one from Combo 1 as the reference and one from Combo 2 as the target, shown in Figure~\ref{fig:city_1_target_model} (the XML representation of both models can be found at \url{https://github.com/ionmatei/ambiguity-detection/tree/main/city-1}. We then apply MBD, with the resulting minimum diagnosis shown in Figure~\ref{fig:city_1_minimum_diagnosis}. The diagnosis identifies the target-model gateways {\tt Check Inclusion Eligibility} and {\tt Check Health Guidance Acceptance} as the main decision points that can explain the observed behavioral differences.

We then use the diagnosis result in the ambiguity detection step. This step links the diagnosed gateways to the policy paragraphs that may have caused the divergent interpretations. For City 1, it yields two ambiguities, with the first ambiguity (AMB-1) associated with the paragraph in Listing~\ref{lst:city_1_amb1}.

\begin{lstlisting}[
  style=json,
  numbers=none,
  basicstyle=\ttfamily\scriptsize,
  breaklines=true,
  breakautoindent=false,
  breakindent=0pt,
  columns=fullflexible,  
  caption={AMB 1, Paragraph in the original City 1 policy description},
  label={lst:city_1_amb1},
  literate={≥}{{$\ge$}}1 
           {–}{{-}}1 
           {㎡}{{$\text{m}^2$}}1 
           {±}{{$\pm$}}1
]
Selection method: Persons who are visiting medical institutions for diabetes and whose renal function is impaired. Judgment criteria based on health checkup results among those who underwent health checkups: fasting blood glucose ≥126 mg/dL or HbA1c ≥6.5%, and eGFR 30–60 mL/min/1.73㎡ or urinary protein ± or higher. Selection target criteria – judgment criteria based on receipts: Persons whose main disease is diabetes caused by lifestyle related diseases. Other judgment criteria: Persons deemed necessary by a physician.  
\end{lstlisting}

The ambiguity report states that this text does not clearly define the condition evaluated at {\tt gateway\_3}, which appears as \texttt{Check Clinical Eligibility} in the reference model and \texttt{Check Inclusion Eligibility} in the target model. In the reference model, {\tt gateway\_3} is interpreted strictly: a patient must satisfy both the receipt-based diabetes criterion and the laboratory thresholds. In the target model, {\tt gateway\_3} is interpreted more broadly: a patient qualifies if either the laboratory thresholds or the receipt-based diabetes criterion is satisfied. This difference changes which patients proceed to notification and therefore produces different simulation traces. The second ambiguity (AMB-2) is associated with the paragraph in Listing~\ref{lst:city_1_amb2}.

\begin{lstlisting}[
  style=json,
  numbers=none,
  basicstyle=\ttfamily\scriptsize,
  breaklines=true,
  breakautoindent=false,
  breakindent=0pt,  
  breakatwhitespace=false,
  breakautoindent=true,
  columns=fullflexible,
  keepspaces=true,
  escapeinside={(*@}{@*)}, 
  caption={AMB 2, Paragraph in the original City 1 policy description},
  label={lst:city_1_amb2},
  literate={≥}{{$\ge$}}1 
           {–}{{-}}1 
           {㎡}{{$\text{m}^2$}}1 
           {±}{{$\pm$}}1
]
Application for participation: Applicants wishing to participate submit a consent form to their primary care physician. The primary care physician submits the consent form and an instruction/recommendation form to the ward.  
Implementation – Content of implementation: Public health nurses and registered dietitians at cooperating medical institutions in the ward that implement health guidance provide health guidance through individual interviews once a month, six times in total.  
\end{lstlisting}

For AMB-2, the ambiguity report states that the text does not clearly define what {\tt Health\_Guidance == 1} means at {\tt gateway\_4}, labeled \texttt{Check Health Guidance Acceptance}. In the reference model, {\tt gateway\_4} is interpreted narrowly: acceptance means that an already eligible patient has consented and is expected to proceed with health guidance. In the target model, {\tt gateway\_4} is interpreted more broadly: submitting the consent form is treated as acceptance, even for patients admitted under the more permissive eligibility logic. This difference changes which patients proceed to health-guidance sessions and explains the different simulation traces. The full ambiguity report for City 1 can be found at \url{https://github.com/ionmatei/ambiguity-detection/blob/main/city-1/city_1_ambiguity_report.pdf}. The final step repairs the policy text to remove the identified ambiguities, with the repair results shown in Listings \ref{lst:city_1_amb1_revision} and~\ref{lst:city_1_amb2_revision}. For AMB-1, the repair makes the clinical eligibility logic explicit by stating that a person must satisfy both a diabetes-related condition and a reduced-renal-function condition, with OR relations within each group. For AMB-2, the repair clarifies that acceptance of health guidance requires both clinical eligibility and submission of the consent form through the primary care physician. Together, these revisions align the narrative more closely with the intended executable logic, reduce ambiguity at the main eligibility and participation gateways, and make the resulting BPMN behavior more consistent under repeated generation. The full repair report that include supplemental material content used for as rational for repairs can be found at \url{https://github.com/ionmatei/ambiguity-detection/blob/main/city-1/city_1_repair_report.pdf}.

\begin{lstlisting}[
  style=json,
  numbers=none,
  basicstyle=\ttfamily\scriptsize,
  breaklines=true,
  breakatwhitespace=false,
  breakautoindent=false,
  breakindent=0pt,  
  breakautoindent=true,
  columns=fullflexible,
  keepspaces=true,
  escapeinside={(*@}{@*)},
  caption={AMB 1, City 1, Revised paragraph and rationale},
  label={lst:city_1_amb1_revision},
  literate={≥}{{$\ge$}}1
           {–}{{-}}1
           {㎡}{{$\text{m}^2$}}1
           {±}{{$\pm$}}1
]
(*@\textbf{revised\_process\_narrative\_excerpt:}@*)
"Selection method: Persons who are visiting medical institutions for diabetes and whose renal function is impaired. Judgment criteria based on health checkup results among those who underwent health checkups: persons who (meet at least one of the following diabetes conditions: (fasting blood glucose ≥126 mg/dL OR HbA1c ≥6.5% OR are currently receiving medical treatment for diabetes OR have a history of diabetes)) AND who (meet at least one of the following reduced renal function conditions: eGFR <45 mL/min/1.73㎡ OR urinary protein (±) or higher OR confirmation of microalbuminuria). Selection target criteria – judgment criteria based on receipts: Persons whose main disease is diabetes caused by lifestyle-related diseases. Other judgment criteria: Persons deemed necessary by a physician,"
(*@\textbf{rationale:}@*)
The revision makes explicit that clinical eligibility requires BOTH (having diabetes) AND (having reduced kidney function), with OR relationships inside each group, matching the supplemental algorithmic structure. The previous phrasing could be read as tying receipts-based diabetes and physician judgment as alternative independent paths; the new text confines eligibility to the combined lab/clinical criteria expressed as ((diabetes conditions) AND (renal conditions)) in line with the final expression `( (Fasting_Blood_Glucose >= 126 OR HbA1c >=6.5) OR Diabetes_Under_Treatment == 1 OR Diabetes_History == 1 ) AND ( eGFR < 45.0 OR Urinary_Protein >= 0 OR Urine_Albumin == 1 )`.
\end{lstlisting}

\begin{lstlisting}[
  style=json,
  numbers=none,
  basicstyle=\ttfamily\scriptsize,
  breaklines=true,
  breakatwhitespace=false,
  breakautoindent=false,
  breakindent=0pt,  
  breakautoindent=true,
  columns=fullflexible,
  keepspaces=true,
  escapeinside={(*@}{@*)},
  caption={AMB 2, City 1, Revised paragraph and rationale},
  label={lst:city_1_amb2_revision},
  literate={≥}{{$\ge$}}1
           {–}{{-}}1
           {㎡}{{$\text{m}^2$}}1
           {±}{{$\pm$}}1
]
(*@\textbf{revised\_process\_narrative\_excerpt:}@*)
"Application for participation: Applicants wishing to participate submit a consent form to their primary care physician. The primary care physician submits the consent form and an instruction/recommendation form to the ward. A person is regarded as having accepted health guidance (Health_Guidance==1) if and only if (they have been deemed eligible according to the program's clinical eligibility criteria AND they submit this consent form through their primary care physician)
Implementation – Content of implementation: Public health nurses and registered dietitians at cooperating medical institutions in the ward that implement health guidance provide health guidance through individual interviews once a month, six times in total."
(*@\textbf{rationale:}@*)
The revision explicitly links the process variable `Health_Guidance == 1` to two conditions: prior eligibility based on clinical criteria and the act of submitting the consent form via the primary care physician. This removes ambiguity about whether consent alone, or consent plus eligibility, constitutes acceptance, and anchors gateway behavior to the operational workflow description where consent is checked after eligibility assessment and notification.
\end{lstlisting}

\begin{figure}[htp!]
    \centering
    \includegraphics[width=0.67\textwidth]{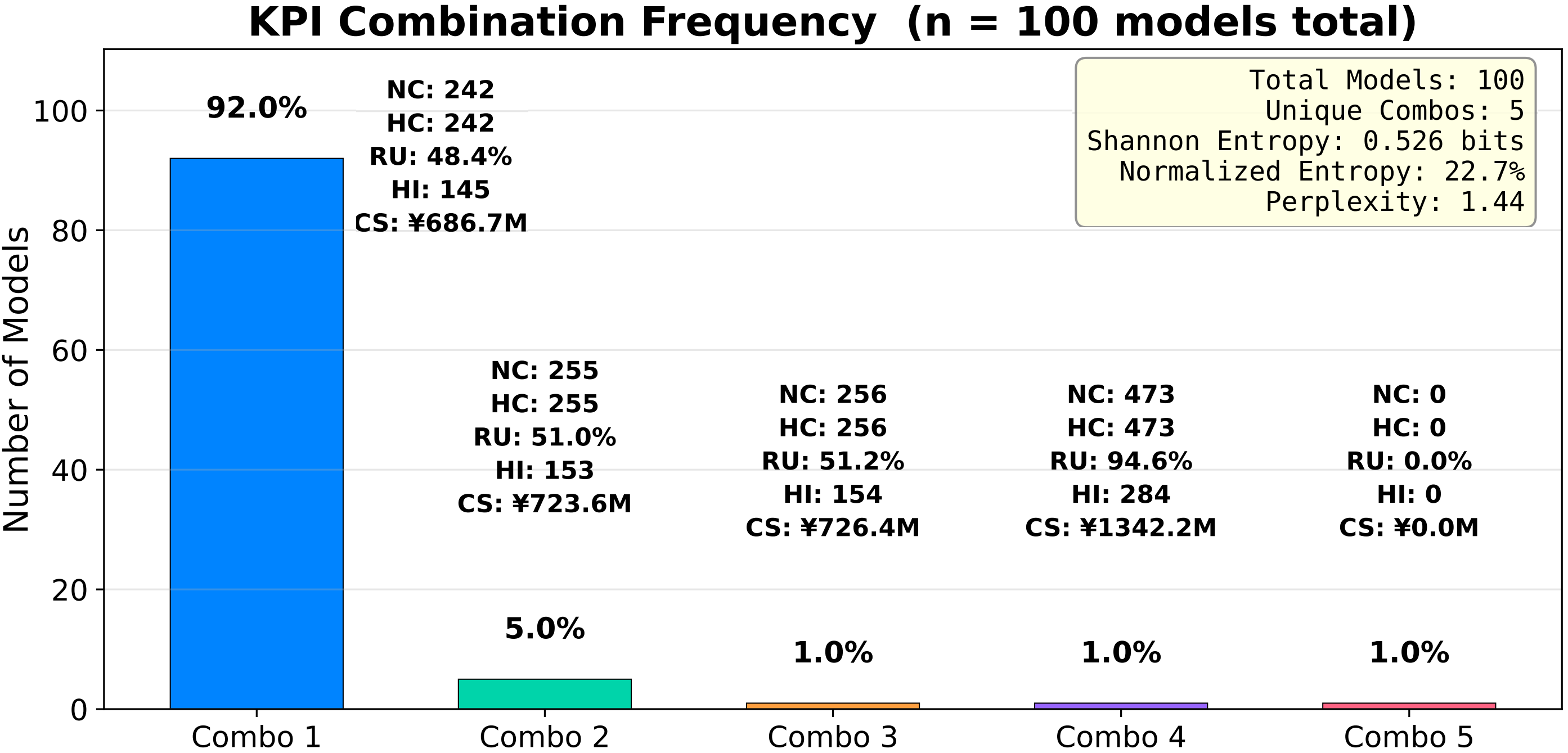}
    \caption{City 1, repaired process description --- Distribution of all five KPIs across 100 generated models.}
    \label{fig:city_1_kpi_uncertainty_repaired}
\end{figure}

We then re-ran the generation process using the repaired policy text, and the resulting histogram is shown in Figure~\ref{fig:city_1_kpi_uncertainty_repaired}. More than 90\% of the generated models now produce the same outcome. Under the entropy categories defined earlier, this corresponds to very high generation consistency after repair and indicates that the clarified process logic substantially reduced behavioral variability.

\subsection{City 2}

\begin{figure}[htp!]
    \centering
    \includegraphics[width=0.7\textwidth]{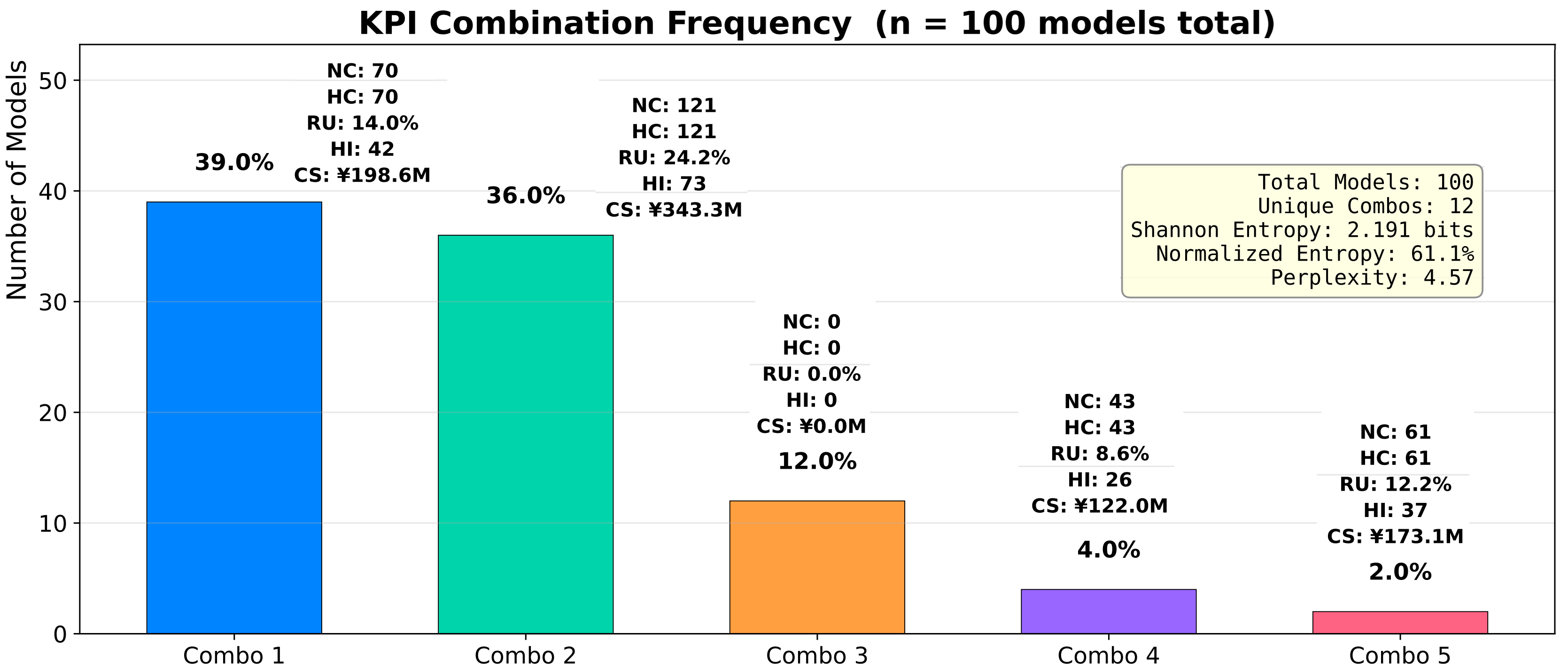}
    \caption{City 2, original process description --- Distribution of all five KPIs across 100 generated models, showing the five most frequent KPI combinations.}
    \label{fig:city_2_KPI_uncertainty_original}
\end{figure}

\begin{figure}[htp!]
    \centering
    \includegraphics[width=0.7\textwidth]{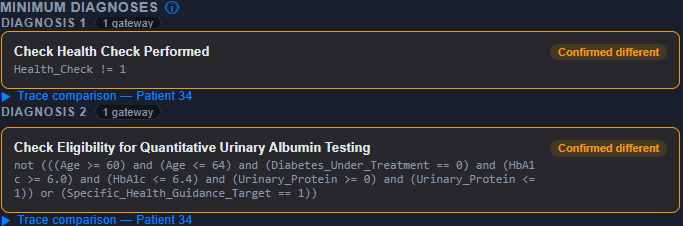}
    \caption{City 2 -- Minimum diagnosis set identifying target-model components that explain the simulation differences.}
    \label{fig:city_2_minimum_diagnosis}
\end{figure}

The City 2 policy (\url{https://github.com/ionmatei/ambiguity-detection/blob/main/city-2/city_2_raw_policy.pdf}) is more operational and clinically specific than City 1. It defines explicit eligibility conditions, downstream screening logic, exclusion criteria, and concrete follow-up actions. Using the same pipeline, we generated 100 BPMN models and computed the KPI distribution shown in Figure~\ref{fig:city_2_KPI_uncertainty_original}. The figure again shows substantial output variability. For the MBD step, we selected a model from Combo 2 as the reference and a model from Combo 1 as the target (both models can be found at \url{https://github.com/ionmatei/ambiguity-detection/tree/main/city-2} with the resulting diagnosis is shown in Figure~\ref{fig:city_2_minimum_diagnosis}.

City 2 is operationally more explicit than City 1 and therefore yields a more branched BPMN structure, with detailed screening, exclusion, and follow-up logic. Under repeated generation, it also exhibits substantial KPI variability, making it a useful second test case for ambiguity detection and repair.

% \begin{figure}[htp!]
%     \centering
%     \includegraphics[width=0.75\textwidth]{figures/city_2_target.png}
%     \caption{City 2: Target model (minimum diagnosis set highlighted in red).}
%     \label{fig:city_2_target_model}
% \end{figure}

Using this diagnosis result, the ambiguity report for City 2 identifies four sources of divergence between the reference and target BPMN models. The first found ambiguity is shown in AMB-1 (Listing~\ref{lst:city_2_amb1}), and concerns the gateway {\tt Check Eligibility for Quantitative Albumin Testing} in the target model. The policy text is unclear about whether eligibility should be based only on Category A, as in the reference model, or on the full condition Category A OR Category B, as in the target model. The full list of ambiguities for City 2 can be found at \url{https://github.com/ionmatei/ambiguity-detection/blob/main/city-2/city_2_ambiguity_report.pdf}.

\begin{lstlisting}[
  style=json,
  numbers=none,
  basicstyle=\ttfamily\scriptsize,
  breaklines=true,
  breakatwhitespace=false,
  breakautoindent=false,
  breakindent=0pt,  
  breakautoindent=true,
  columns=fullflexible,
  keepspaces=true,
  escapeinside={(*@}{@*)},
  caption={AMB 1, Paragraph in the original City 2 policy description},
  label={lst:city_2_amb1},
  literate={≥}{{$\ge$}}1 
           {–}{{-}}1 
           {㎡}{{$\text{m}^2$}}1 
           {±}{{$\pm$}}1
           {①}{{\textcircled{1}}}1
           {−}{{-}}1
]
An individual is eligible for quantitative urinary albumin testing if, and only if, at least one of the following two categories applies: - Category ①-A: The individual (a) is 60 to 64 years of age at the end of the fiscal year AND (b) is not receiving diabetes medication AND (c) has R5 health checkup results showing HbA1c in the range 6.0% to 6.4% AND (d) has urinary protein in the range from (−) to (+). In logical form: (Age is between 60 and 64 at the end of the fiscal year) AND (not receiving diabetes medication) AND (HbA1c is between 6.0% and 6.4% in the R5 health checkup) AND (urinary protein is between (−) and (+)). - Category ①-B: The individual is designated in R5 as a target of the diabetes medical consultation recommendation program. In logical form: (Designated in R5 as a target of the diabetes medical consultation recommendation program). The overall eligibility condition for ① is: (Category ①-A) OR (Category ①-B).
\end{lstlisting}

The ambiguity report was then used in the repair step. The results for repairing AMB-1 are shown in Listing \ref{lst:city_2_amb1_revision}, with the full repair report found at \url{https://github.com/ionmatei/ambiguity-detection/blob/main/city-2/city_2_repair_report.pdf}. The first two ambiguities (AMB-1 and AMB-2) are resolved by making the quantitative urinary albumin testing pathway explicit, with eligibility defined as Category \ding{172}-A OR Category \ding{172}-B. Individuals who satisfy neither condition are excluded, and the testing workflow terminates. The third ambiguity (AMB-3) is resolved by clarifying that the sequence ``eligible for testing $\rightarrow$ tested $\rightarrow$ microalbuminuria'' is only one sufficient path to Renal Health Screening interview eligibility, not the only one. This preserves physician-determined criteria as an independent basis for participation. The forth ambiguity (AMB-4) is resolved by clarifying who receives outreach. Telephone contact and ticket issuance are tied specifically to high-risk individuals identified from R5 health check data. Missing qualifying health-check information is also made an explicit reason for exclusion from the testing pathway. Together, these revisions make the gateway logic, termination behavior, and downstream eligibility conditions much clearer and lead to more consistent generated BPMN behavior.

\begin{lstlisting}[
  style=json,
  numbers=none,
  basicstyle=\ttfamily\scriptsize,
  breaklines=true,
  breakatwhitespace=false,
  breakautoindent=false,
  breakindent=0pt,  
  breakautoindent=true,
  columns=fullflexible,
  keepspaces=true,
  escapeinside={(*@}{@*)},
  caption={AMB 1, City 2, Revised paragraph and rationale},
  label={lst:city_2_amb1_revision},
  literate={≥}{{$\ge$}}1
           {–}{{-}}1
           {㎡}{{$\text{m}^2$}}1
           {±}{{$\pm$}}1
           {①}{{\ding{172}}}1
]
(*@\textbf{revised\_process\_narrative\_excerpt:}@*)
An individual is eligible for quantitative urinary albumin testing if, and only if, at least one of the following two categories applies, and individuals who do not satisfy either category are not eligible and shall exit this testing-related workflow at this point:
- Category ①-A: The individual (a) is 60 to 64 years of age at the end of the fiscal year AND (b) is not receiving diabetes medication AND (c) has R5 health checkup results showing HbA1c in the range 6.0% to 6.4% AND (d) has urinary protein in the range from (-) to (+). In logical form:
(Age is between 60 and 64 at the end of the fiscal year) AND (not receiving diabetes medication) AND (HbA1c is between 6.0% and 6.4% in the R5 health checkup) AND (urinary protein is between (-) and (+)).
- Category ①-B: The individual is designated in R5 as a target of the diabetes medical consultation recommendation program, based on the program's diabetes and kidney-function criteria. In logical form:
(Designated in R5 as a target of the diabetes medical consultation recommendation program).
The overall eligibility condition for ① is:
((Category ①-A) OR (Category ①-B)); individuals for whom NOT((Category ①-A) OR (Category ①-B)) holds are not eligible for quantitative urinary albumin testing and are excluded from further processing related to this testing.",
(*@\textbf{rationale:}@*)
"The revision makes explicit that the logical condition for eligibility is exactly (Category ①-A OR Category ①-B), and that failing both categories results in termination of the quantitative-test-related workflow. This selects the \"full OR\" interpretation and removes the possibility of implementing only Category ①-A. The explicit NOT((A) OR (B)) clause aligns the narrative with a gateway whose negative branch ends processing for non-eligible individuals."
\end{lstlisting}

% \begin{figure}[htp!]
%     \centering
%     \includegraphics[width=0.75\textwidth]{figures/city_2_amb1_repaired.png}
%     \caption{City 2: Repaired AMB 1.}
%     \label{fig:city_2_amb1_repaired}
% \end{figure}

\begin{figure}[htp!]
    \centering
    \includegraphics[width=0.7\textwidth]{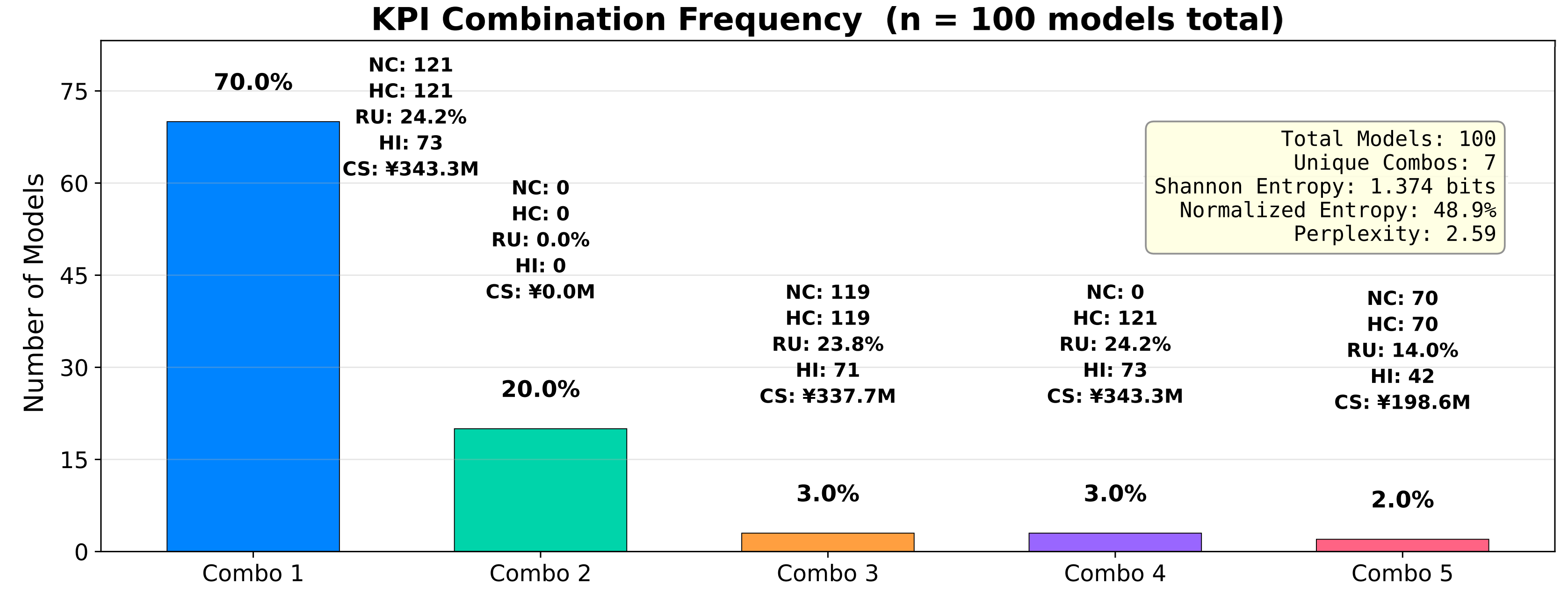}
    \caption{City 2, repaired process description --- Distribution of all five KPIs across 100 generated models.}
    \label{fig:city_2_kpi_uncertainty_repaired}
\end{figure}

After re-running the generation process, we obtained the histogram in Figure~\ref{fig:city_2_kpi_uncertainty_repaired}, with 70\% of the generated models now producing the same outcome. Under the entropy categories defined earlier, this corresponds to high generation consistency after repair and indicates that the clarified process logic reduced behavioral variability.

\section{Comparison with the State of the Art and Discussion}
\label{sec:Comparison}

Our work relates to three research areas: ambiguity detection in natural-language specifications, ambiguity repair, and automated process-model generation. Its distinguishing feature is that ambiguity is not treated as a purely linguistic property. Instead, we detect and explain ambiguity through its observable effect on the behavior of executable BPMN models under simulation.

Ambiguity has long been recognized as a major source of defects in requirements engineering. Prior work defined ambiguity from dictionary, linguistic, and software-engineering perspectives and surveyed techniques and tools for avoiding and detecting ambiguity in natural-language requirements~\cite{berry2003ambiguity}. Rule-based tools such as QuARS detect potential ambiguity-related defects through lexical and syntactic analysis~\cite{lami2005quars}, while standards such as ISO/IEC/IEEE~29148 emphasize unambiguity as a core requirement-quality property~\cite{iso29148_2018}. More recent work uses machine learning and LLMs to detect and explain ambiguity, extending beyond earlier rule-based approaches that often relied on ambiguous keywords, key phrases, or heuristics~\cite{bashir2025requirementsambiguityllm}. However, these methods remain primarily text-centric and typically do not directly evaluate whether alternative readings lead to materially different executable behavior. 
In contrast, our framework flags ambiguity as actionable only when it is supported by behavioral evidence, namely divergent KPI outcomes produced by executable BPMN models generated from the same source text. One line of prior work addresses ambiguity prevention or reduction through controlled natural language, such as Attempto Controlled English, which rewrites requirements/specifications into a more precise, machine-processable form~\cite{fuchs1999attempto}. 
These approaches can be effective, but many require substantial human involvement or impose strong upfront language restrictions. Other methods use formal artifacts, examples, or human-in-the-loop feedback to guide refinement, but their applicability often depends on the availability of supporting annotations, domain knowledge, or expert validation~\cite{beg2025leveragingllmsformalsoftware,YADAV202185}. LLM-based rewriting can generate plausible clarifications, but without a principled basis for choosing among interpretations, it may introduce edits that are linguistically fluent yet weakly justified. Our repair method addresses this limitation by grounding revision in diagnosed behavioral discrepancies, mappings from diagnosed gateway logic to verbatim narrative segments, and authoritative supplemental evidence. As a result, the revised text is targeted at ambiguities shown to alter executable behavior. 

Our work is also related to recent research on automated BPMN generation and ambiguity-aware process modeling~\cite{kourani2024processmodelingllm,vanDerAa2019text2process}. These studies have shown that LLMs can extract process elements and support iterative model refinement, sometimes with human feedback or ambiguity-aware prompting. Their main focus, however, is improving generation quality. Our focus is different. We address the case in which repeated generations from the same specification produce multiple structurally valid but behaviorally different models. In that setting, the key question is not only how to generate a BPMN model, but how to determine whether the source text supports a stable executable interpretation. We answer this question through a diagnosis-driven loop that connects simulation outputs, gateway-level discrepancies, narrative ambiguity, and evidence-based repair.

The closest methodological distinction of our work is therefore the role of executable evidence. Rather than inferring ambiguity solely from wording, we use simulation-output divergence to decide which ambiguities matter, model-based diagnosis to localize the responsible decision logic, and regeneration plus re-simulation to test whether the repair improves behavioral consistency. This closed-loop validation structure provides a level of behavioral grounding and traceability that text-only approaches do not offer. Our method has two main limitations. First, it can detect only ambiguities that affect the monitored KPIs under the sampled input population. Ambiguities that do not change those outputs, or that arise outside the explored input region, may remain undetected. Second, the localization and repair stages still rely on LLM prompting and on the completeness of the supporting evidence. Therefore, the quality of the final revision depends on both prompt design and the quality of the external material used to resolve competing interpretations.

\section{Conclusion}

We presented a diagnosis-driven framework for validating and repairing LLM-generated executable BPMN models when no ground-truth BPMN is available. The method uses repeated simulation to detect behavioral inconsistency, MBD to localize divergence to gateways, and evidence-based text repair to remove the responsible ambiguity from the source narrative. In two diabetic nephropathy policy case studies, the repaired specifications produced substantially more concentrated KPI distributions under regeneration. The main contribution is a closed validation loop from natural-language specification to executable model, simulated behavior, diagnosis, and repaired specification.

\bibliographystyle{plain}
\bibliography{references}  

\end{document}